\documentclass{bioinfo_arxiv}
\copyrightyear{2011}
\pubyear{2011}

\newcommand{\by}{\mbox{\boldmath $y$}}

\newcommand{\bbeta}{\mbox{\boldmath $\beta$}}

 \usepackage{graphicx}

\usepackage{url}

\begin{document}
\firstpage{1}

\title[Clustering Time Course Gene-Expression Profiles]
{Clustering of time-course gene expression profiles using normal mixture models with AR(1) random effects}
\author[Wang \textit{et~al.}]{
K. Wang\,$^{\rm a}$, 
S.K. Ng\,$^{\rm b}$,
G.J. McLachlan\,$^{\rm c}$\footnote{to whom correspondence should be
addressed}}
\address{$^{\rm a}$Department of Mathematics, University of
Queensland, Brisbane, QLD 4072, Australia, 
$^{\rm b}$School of Medicine, Griffith University (Logan Campus), 
Meadowbrook, QLD 4131, Australia,
$^{\rm c}$Institute for Molecular Bioscience, University of
Queensland, Brisbane, QLD 4072, Australia.} 
\maketitle

\begin{abstract}

\section{Motivation:}
Time-course gene expression data such as yeast cell cycle data may 
be periodically expressed. 
To cluster such data, currently used Fourier series approximations of 
periodic gene expressions have been found not to be sufficiently
adequate to model the complexity of the time-course data, 
partly due to their ignoring the dependence between the expression 
measurements over time and the correlation among gene expression profiles. 
We further investigate the advantages and limitations of available models 
in the literature and propose a new mixture model with AR(1) random effects 
for the clustering of time-course gene-expression profiles. 
Some simulations and real examples are given to demonstrate the usefulness
of the proposed models.

\section{Results:}
We illustrate the applicability of our new model using synthetic and real 
time-course datasets. We show that our model outperforms existing models 
to provide more reliable and robust clustering of time-course data. 
Our model provides superior results when genetic profiles are correlated. 
It also gives comparable results when the correlation between the gene 
profiles is weak. 
In the applications to real time-course data, relevant clusters of 
co-regulated genes are obtained, which are supported by gene-function 
annotation databases. 

\section{Availability:}An R-program is available on request from the 
corresponding author.

\section{Contact:} \href{g.mclachlan@uq.edu.au}{g.mclachlan@uq.edu.au}

\section{Supplementary Information:}
\verb+http://www.maths.uq.edu.au/+ \\
\verb+~gjm/bioinf_2011_supp.pdf+.
\end{abstract}

\section{Introduction}

DNA microarray analysis has emerged as a leading technology to enhance 
our understanding of gene regulation and function in cellular mechanism 
controls on a genomic scale. This technology has advanced to unravel the 
genetic machinery of biological rhythms by collecting massive gene-expression 
data in a time course. 
Time-course gene expression data such as yeast cell 
cycle data \citep{wichert04} appear to be periodically expressed. 
To associate the profile of gene expression with a physiological function 
of interest, it is crucial to cluster the types of gene expression on the 
basis of their periodic patterns. 
The identification of co-expressed genes
also facilitates the prediction of response to treatment or toxic 
compounds \citep{hafemeister11}. 
Statistical modelling and algorithms play 
a central role in cataloguing dynamic gene-expression profiles.

Various computational models have been developed for gene clustering 
based on cross-sectional microarray data 
(\citealp{gjm02}; \citealp{ram02}; \citealp{fan06}).
Also, considerable attention has been paid to methodological 
derivations for detecting temporal patterns of gene expression in a time 
course based on functional principal component analysis or mixture model 
analysis (\citealp{qin06}; \citealp{xu02}; \citealp{luan03}; 
\citealp{luan04}; \citealp{storey05}; \citealp{hong06}; \citealp{ma06}; 
\citealp{ng06}; \citealp{kim08}; \citealp{booth08}), including the applications
to identify differentially expressed genes over time 
(\citealp{park03}; \citealp{sun11}).
 
Finite mixture models \citep{gjm00} have been widely used to model 
the distributions of a variety of random phenomena. 
Multivariate normality is generally assumed for multivariate data 
of a continuous nature. 
The multivariate normal mixture model is employed to detect different patterns 
in gene-expression profiles. 
However, when the two assumptions that are commonly adopted in practice, 
namely, 
 
(1)	there are no replications on any particular entity specifically 
identified as such and 

(2)	all the observations on the entities are independent of one another, 

\noindent
are violated, multivariate normal mixture models may not be adequate. 
For example, condition (2) will not hold for the clustering of gene profiles, 
since not all the genes are independently distributed, 
and condition (1) will generally not hold either as the gene profiles 
may be measured over time or on technical replicates. 
While this correlated structure can be incorporated into the 
normal mixture model by appropriate specification of the 
component-covariance matrices, it is difficult to fit the model 
under such specifications. 
For example, the M-step may not exist in closed form \citep{gjm04}. 

Accordingly, \cite{ng06} have developed the procedure called 
EMMIX-WIRE ({\bf EM}-based {\bf MIX}ture analysis{\bf Wi}th 
{\bf R}andom Effects) to handle the clustering of correlated data 
that may be replicated. 
They adopted a mixture of linear mixed models 
to specify the correlation structure between the variables 
and to allow for correlations among the observations. 
It also enables covariate information to be incorporated 
into the clustering process \citep{ng06}. 
Proceeding conditionally on the tissue-specific random effects 
as formulated in \cite{ng06}, 
the E- and M-steps can be implemented in closed form. 
In particular, an approximation to the E-step by carrying out 
time-consuming Monte Carlo methods is not required. 
A probabilistic or an outright clustering of the genes into g components 
can be obtained, based on the estimated posterior probabilities of 
component membership given the profile vectors and the 
estimated tissue-specific random effects; see \cite{ng06}. 

Fourier series approximations
have been used to model periodic gene expression, 
leading to the detection of periodic signals in various organisms 
including yeast and human cells (\citealp{spellman98}; 
\citealp{wichert04}; \citealp{kim06}). 
If the genes studied are periodically regulated, 
their time-dependent expression can be accurately approximated by 
a Fourier series approximation \citep{spellman98}. 
A general form of the $k$th order Fourier series expansion is given as 
\begin{equation} 
g_k(t) = a_0+\sum_{j=1}^k[a_j {\rm cos}(2\pi jt/\omega)
+b_j {\rm sin}(2\pi jt/\omega],
\label{eq:1}
\end{equation}
where $a_0$ is the average value of $g_k(t)$. 
The other coefficients $a_k$ and $b_k$ are the amplitude coefficients 
that determine the times at which the gene achieves peak and trough expression 
levels, respectively, and $\omega$ is the period of the signal of 
gene expression. 
While the time-dependent expression value of a gene can be adequately 
modelled by a Fourier series approximation of the first three orders 
\citep{kim08}, recent results (\citealp{kim08}; \citealp{ng06}) demonstrate 
that the first-order Fourier series approximation is sufficient to provide 
good results in terms of clustering the time-course data into meaningful 
functional groups. 
Alternatively, the likelihood ratio test may be used to determine the order 
of the Fourier series approximation within the nested regression models. 

The EMMIX-WIRE model 
of \cite{ng06} is developed primarily for clustering genes from 
general microarray experimental designs. 
On the other hand, \cite{kim08} focus specifically on clustering periodic 
gene profiles and propose a special covariance structure 
to incorporate the correlation between observations at different time points. 
They also review current methods and compare their method with that of 
\cite{ng06}. 
More recently, \cite{scharl10} use integrated autoregressive (AR) models 
to create cluster centers in their simulation study of mixtures 
of regression models for time-course gene expression data 
through the new version of software FlexMix of \cite{leisch04}. 
\cite{wang10} propose mixtures of multivariate linear mixed models 
with autoregressive errors to analyse longitudinal data. 
In this paper, we propose a new EMMIX-WIRE normal mixture regression model 
with AR(1) random effects for the clustering of time-course data. 
In particular, the model accounts for the correlation among gene profiles 
and models the dependence between expressions over time 
via AR(1) random effects.

The paper is organized as follow: 
Section 2 presents the development of the extension of the EMMIX-WIRE model 
to incorporate AR(1) random effects which are fitted under the EM framework.
We conduct a simulation study and the data analysis with two real yeast cell
data in Section 3. 
In the last section, some discussion is provided. 
The technical details of the derivations are provided in the 
Supplementary Information.

\section{	EMMIX-WIRE MODEL WITH AR(1) RANDOM EFFECTS}

We let $X$ denote the design matrix and $\beta$ 
the associated vector of regression 
coefficients for the fixed effects. 
In the specification of the mixture of mixed linear components
as adopted by \cite{ng06}, the vector $\by_j$ for the $j$th gene 
conditional on its membership of the
$h$th component of the mixture is expressed as
\begin{equation} 
\by_j = X \bbeta_h + Z_1 u_{jh} + Z_2 v_h + \epsilon_{jh} \quad (j=1,\ldots,n),
\label{eq:2}
\end{equation}
where $\beta_h$ is a $(2k+1)$ vector containing unknown parameters 
$a_0, a_1, \dots, a_k, b_1, \dots, b_k$; see (1),
$u_{jh}=(u_{jh1},\dots,u_{jhm})^T$ and $v_h=(v_{h1},\dots,v_{hm})^T$ are the
random effects, where $m$ is the number of time points. In (\ref{eq:2}),
$Z_1$ and $Z_2$ are $m \times m$ identity matrices. 
Without loss of generality,
we assume $\epsilon_{jh}$ and $v_h$ to be independent 
and normally distributed, 
$N(0,\Omega)$ and $N(0,D)$, independent of $u_{jh}$. 
To further account 
for the time dependent random gene effects, a first-order autoregressive 
correlation structure is adopted for the gene profiles, so that $u_{jh}$ 
follows a $N(0, \theta^2A(\rho))$ distribution, where 
\begin{equation} 
A(\rho) = { \frac{1}{1-\rho^2 }}  \left( \begin{array}{cccc}
1     & \rho   & \dots  & \rho^{m-1} \\
\rho  & 1      & \dots  & \rho^{m-2} \\
\vdots & \vdots  & \vdots  & \vdots      \\
\rho^{m-1} & \rho^{m-2} & \dots & 1 \end{array} \right).
\label{eq:3}
\end{equation}
The inverse of $A(\rho)$ can be expressed as
\begin{equation}
A(\rho)^{-1}=(1+\rho^2)I-\rho J-\rho^2 K,
\label{eq:3a}
\end{equation}
and
\begin{equation} 
{\rm trace} \left(  {\frac{\partial{A(\rho)^{-1}}}{\partial{\rho}}} 
A(\rho) \right )=-2\rho/(1-\rho^2),
\label{eq:4}
\end{equation}
where $I$, $J$ and $K$ are all $m \times m$ matrices. 
Specifically, $I$ is the identity matrix; $J$ has its sub-diagonal entries 
ones and zeros elsewhere, and $K$ takes on the value 1 at the first and 
last element of its principal diagonal and zeros elsewhere. 
The expressions (\ref{eq:3a}) and (\ref{eq:4}) are needed in the
derivation of the maximum likelihood estimates of the parameters.

The assumptions 
(\ref{eq:2}) and (\ref{eq:3}) imply that our new model
assumes an auto-correlation covariance structure under which 
measurements at each time point have a larger variance compared 
to the model of Kim et al.\ (2008) under an AR(1) auto-correlation 
residual structure. 

In the context of mixture models, we consider the $g$-component mixture 
with probability density function (pdf) as 
\begin{equation} 
f(y\mid \Psi)=\sum_{h=1}^gp_hf_h(y_j\mid \beta_h,\Omega_h,\theta_h^2,A_h,D_h),
\label{eq:5}
\end{equation}
where $f_h$ is the component-pdf of the multivariate normal distribution 
with mean vector $X_h\beta_h$ and covariance matrix  
$$\theta_h^2Z_1A_hZ_1^T+Z_2D_hZ_2^T+\Omega_h.$$
The vector of unknown parameters is denoted by $\Psi$ and can be
estimated by maximum likelihood via the EM algorithm. 

In the EM framework adopted here, the observed data vector  
$y=(y_1,y_2,\dots,y_n)^T$ is augmented by the unobservable component labels, 
$z_1, z_2, \dots, z_n$ of $y_1,y_2, \dots, y_n$ , where $z_j$ is the 
$g$-dimensional vector with $h$th element $z_{jh}$, which is equal to 1 if 
$y_j$ comes from the $h$th component of the mixture, and is zero otherwise. 
These unobservable values are considered to be missing data and are included 
in the so-called complete-data vector. Finally, we take the random effect 
vectors $u_{jh}$ and $v_h$ $(j=1,\ldots,n; \,h=1,\ldots,g)$, to be missing 
and include them too in the complete-data vector. 
Now the so called complete-data log-likelihood $l_c$ 
is the sum of four terms 
$l_c = l_1 + l_2 + l_3 + l_4$, where
\begin{equation}
l_1=\sum_{h=1}^g{\sum_{j=1}^n z_{jh} log(p_h)}
\label{eq:6}
\end{equation}
is the logarithm of the probability of the component labels $z_{jh}$, 
and where $l_2$ is the logarithm of the density function of $y$ 
conditional on $u_{jh}, v_h$, and $z_{jh}$=1, and  $l_3$ and $l_4$ is 
the logarithm of the density function of $u$ and $v$, respectively,    
given $z_{jh}$=1,
\begin{equation} 
l_2=-{\textstyle\frac{1}{2}}
\sum_{h=1}^g{\sum_{j=1}^n z_{jh} \left 
( mlog(2\pi)+log|\Omega_h|+\epsilon_{jh}^T\Omega_h^{-1}\epsilon_{jh}\right )},
\label{eq:7}
\end{equation}
\begin{equation} 
l_3=-{\textstyle\frac{1}{2}}
\sum_{h=1}^g{\sum_{j=1}^n z_{jh} \left ( mlog(2\pi 
\theta_h^2)+log|A_h|+\theta_h^{-2}u_{jh}^T A_h^{-1}u_{jh}\right )},
\label{eq:8}
\end{equation}
\begin{equation} 
l_4=-{\textstyle\frac{1}{2}}
\sum_{h=1}^g{(\sum_{j=1}^n z_{jh}) 
\left ( mlog(2\pi)+log|D_h|+v_{h}^T D_h^{-1}v_{h}\right )},
\label{eq:9}
\end{equation}
where   $$\epsilon_{jh}=y_j-X \beta_h-Z_1 u_{jh} - Z_2 v_h.$$        
	   
To maximize the complete-data log likelihood $l_c$, 
the above decomposition 
implies that each of $l_1$, $l_2$, $l_3$, and $l_4$ can be maximized 
separately.
The EM algorithm proceeds iteratively
until the difference between successive values of the log likelihood is
less than some specified threshold.
All major derivations are given in the Supplementary Information.

\begin{table}[!t] 
\processtable{Bias and standard deviation in brackets from 1000 simulated 
data points (generated from new EMMIX-WIRE (EM-W) model with $\theta_h^2$ equal 
to 0.5) \label{Tab:1}}
{\begin{tabular}{lllllll}\toprule
&\multicolumn{2}{c}{First component}&\multicolumn{2}{c}{Second component}
&\multicolumn{2}{c}{Third component}\\
\cline{2-7}
Parameters&EM-W&Kim&EM-W&Kim&EM-W&Kim\\\midrule
$p$(0.585,&-0.002 & 0.016 &-0.009 & -0.001 & 0.011 &-0.015 \\
0.1,0.315)&(0.045)&(0.052) &(0.033) &(0.029)&(0.051) &(0.051)\\
$a_0$(0.3,&0.002 &0.008 &-0.006 & -0.036 &-0.003 &-0.009\\
1,0.2)&(0.135)&(0.137)& (0.175) &(0.186) &(0.186) &(0.182)\\
$a_1$(0.03,&-0.001 &-0.018 &0.024 &0.004 &0.004 &-0.001\\
1,0.02)&(0.119)&(0.124)&(0.272) &(0.160)&(0.175) &(0.152)\\
$b_1$(0.06,&0.009 & -0.015 &-0.164 & 0.031 &0.027 &0.008\\
0.9,0.01)&(0.119) &(0.132) &(0.223)&(0.160)&(0.149)&(0.183)\\
\hline
$\theta^2$(0.5,& 0.055& 1.543 & 0.089 &1.346 &0.110 &1.443 \\
0.5,0.5) &(0.082)&(1.547) & (0.164)& (1.349) &(0.152) &(1.446)\\
$\rho(0.6$ &-0.023 &-0.395 & -0.043  & -0.372 &-0.043 &-0.392 \\
0.6,0.6)&(0.036) &(0.397) & (0.082) &(0.374) &(0.058) &(0.394)\\
\hline
$\sigma^2$(1.0,&0.0171 & &-0.017 && 0.011 & \\
1.0,1.0) &(0.055) & &(0.127) && (0.088) &\\
$d^2$(0.4, & -0.112 & & -0.091 && -0.118 & \\
0.2,0.3) & (0.145) & &(0.102) &&(0.134)&\\
\hline
& & EM-W & & & Kim & \\
\hline
Error rate & &0.036 & & &0.098 &\\
Rand & &0.954 & & & 0.864 & \\
Adjusted & & 0.907 & & & 0.726  & \\\botrule

\end{tabular}}{}
\end{table} 

\section{SIMULATIONS AND APPLICATIONS}
\subsection{Simulation study}

To illustrate the performance of the proposed model, 
we present a simulation study based on synthetic time-course data. 
In the following simulation, we consider an autocorrelation dependence 
for the periodic expressions and compare our model to that of \cite{kim08}. 
Synthetic time-course data from three different parametric models 
(the full model under our new extended EMMIX-WIRE approach (denoted by
EM-W in the tables), 
the extended model of \cite{qin06}, 
and the model of \cite{kim08}), assuming a first-order Fourier series 
of periodicity, are considered in the simulation study. 
Within each model, we consider two different settings of 
$\theta^2$ corresponding to low and high auto-correlation 
among the periodic gene expressions. 
We also assume that $\Omega$ and $D$ are diagonal matrices, 
where the common diagonal elements are represented by 
$\sigma^2$ and $d^2$, respectively.

\begin{table}[!t] 
\processtable{Bias and standard deviation in brackets from 1000 simulated 
data points (generated from new EMMIX-WIRE (EM-W) model 
with $\theta_h^2$ equal to 1.3) \label{Tab:2}}
{\begin{tabular}{lllllll}\toprule
&\multicolumn{2}{c}{First component}&\multicolumn{2}{c}{Second
component}
&\multicolumn{2}{c}{Third component}\\
\cline{2-7}
Parameters&EM-W&Kim&EM-W&Kim&EM-W&Kim\\\midrule
$p$(0.585,    &-0.006 & 0.035  &-0.009  & -0.002 & 0.015  &-0.033 \\
0.1,0.315)  &(0.061)&(0.080) &(0.047) &(0.045) &(0.070) &(0.074)\\
$a_0$(0.3,    &0.001  &0.018   &-0.004  & -0.069 &-0.00   &-0.014\\
1,0.2)      &(0.137)&(0.147) &(0.173) &(0.197) &(0.186) &(0.178)\\
$a_1$(0.03,   &0.010  &-0.062  &0.017   &-0.031  &0.001   &-0.002\\
1,0.02)     &(0.162)&(0.227) &(0.388) &(0.236) &(0.230) &(0.199)\\
$b_1$(0.06,   &0.009  & -0.042 &-0.180  & 0.073  &0.032   &0.009 \\
0.9,0.01)   &(0.124)&(0.166) &(0.235) &(0.188) &(0.163) &(0.213)\\
\hline
$\theta^2$(1.3,& -0.042& 1.671  & -0.030  &1.449    &0.008   &1.549 \\
1.3,1.3)    &(0.097) &(1.677) & (0.223) & (1.460) &(0.153) &(1.556)\\
$\rho$(0.6    &0.009   &-0.249  & -0.001  & -0.228  &0.002   &-0.250 \\
0.6,0.6)    &(0.020) &(0.251) & (0.055) &(0.235)  &(0.025) &(0.252)\\
\hline
$\sigma^2$(1.0,&0.131 & &0.121   && 0.141   &\\
1.0,1.0) &(0.155)   & &(0.219) && (0.186) &\\
$d^2$(0.4, & -0.151   & & -0.124 && -0.160  &\\
0.2,0.3) & (0.172)  & &(0.129) &&(0.168) &\\
\hline
& & EM-W & & & Kim & \\
\hline
Error rate & &0.094  & & & 0.184  & \\
Rand       & &0.881  & & & 0.759  & \\
Adjusted   & &0.760  & & & 0.519  & \\\botrule

\end{tabular}}{}
\end{table} 
 
There are three classes of genes. The periods for each class 
are 6, 10 and 16, respectively. There are 24 measurements at time points 
0, 1, ..., 23, and the first order Fourier expansion is adopted in the 
simulation models. Parameters and simulation results are listed in Tables 1 
to 6. In each table, we summarize the results from 1000 simulated sets of data.
The true values of the parameters and the means of their estimates are given 
in these tables, along with the standard errors in parentheses.
We terminated the EM algorithm iterations when the absolute values of the
relative changes in all estimates between consecutive iterations were
smaller than $0.00001$, with the maximum iteration of 1000. 
For our model, we started from the true partition; for \cite{kim08}, we started
from the true values of parameters. Alternatively, initialization procedures
have been considered for mixtures of regression models with and without
random effects \citep{scharl10}. For the comparison, we consider the 
misclassified error rate, the Rand Index, and the adjusted Rand 
Index \citep{hubert85}, where the latter two assess the degree of agreement 
between the partition and the true clusters of genes. A larger (adjusted) 
Rand Index indicates a higher level of agreement. 

Specifically, we first investigate the performance of our new extended
EMMIX-WIRE model and that of \cite{kim08} when the data are generated 
from the extended EMMIX-WIRE model, in which gene expressions
within a cluster are correlated. 
As listed in Tables 1 and 2, the estimates of the
parameters $p, a_0, a_1, b_1, \theta^2, \rho$, and $\sigma^2$ in the 
proposed model are approximately unbiased, 
except for $d^2$, which is slightly underestimated. 
In contrast, the method of \cite{kim08} 
fails to capture the contributions from gene-specific and
tissue-specific effects on the auto-correlation 
among periodic gene expressions at each time point, 
and thus overestimates the correlation between different
time points for each gene. 
Their method therefore leads to an inferior clustering performance 
in terms of higher error rates and smaller Rand Indices.

\begin{table}[!t] 
\processtable{Bias and standard deviation in brackets from 1000 simulated 
data points (generated from new EMMIX-WIRE (EM-W) model 
with $\theta_h^2$ equal to 0.5 and $d^2$ equal to 0) \label{Tab:3}}
{\begin{tabular}{lllllll}\toprule
&\multicolumn{2}{c}{First component}&\multicolumn{2}{c}{Second
component}
&\multicolumn{2}{c}{Third component}\\
\cline{2-7}
Parameters&EM-W&Kim&EM-W&Kim&EM-W&Kim\\\midrule
$p$(0.585,    &0.001 & 0.008   &-0.001  & -0.003 & -0.001  &-0.005 \\
0.1,0.315)  &(0.009)&(0.012) &(0.008) &(0.008) &(0.010) &(0.011)\\
$a_0$(0.3,  &0.001  &0.008   &-0.001  & -0.018 &0.003   &-0.014\\
1,0.2)      &(0.017)&(0.019) &(0.018) &(0.026) &(0.016) &(0.016)\\
$a_1$(0.03, &-0.002  &-0.023  &-0.001   &-0.005  &0.003   &-0.006\\
1,0.02)     &(0.049)&(0.060) &(0.059) &(0.062) &(0.049) &(0.049)\\
$b_1$(0.06, &-0.001  & -0.014 &0.016  & 0.019  &0.002   &0.004 \\
0.9,0.01)   &(0.026)&(0.031) &(0.033) &(0.038) &(0.032) &(0.033)\\
\hline
$\theta^2$(0.5,&0.071 & 1.162  & 0.081  &1.158    &0.078   &1.159 \\
0.5,0.5)       &(0.081) &(1.162) & (0.119) & (1.160) &(0.090) &(1.159)\\
$\rho$(0.6     &-0.032   &-0.337  & -0.037  & -0.339  &-0.036   &-0.339 \\
0.6,0.6)       &(0.038) &(0.337) & (0.062) &(0.340)  &(0.045) &(0.340)\\
\hline
$\sigma^2$(1.0,&-0.059 & &-0.069   && -0.064   &\\
1.0,1.0)       &(0.068)   & &(0.106) && (0.077) &\\
$d^2$(0,       & 0   & & 0.001 && 0.000  &\\
0,0)           & (0.000)  & &(0.001) &&(0.001) &\\
\hline
& & EM-W & & & Kim & \\
\hline
Error rate & &0.078  & & & 0.081  & \\
Rand       & &0.891  & & & 0.886  & \\
Adjusted   & &0.780  & & & 0.769  & \\\botrule

\end{tabular}}{}
\end{table} 

\begin{table}[!t] 
\processtable{Bias and standard deviation in brackets from 1000 simulated 
data points (generated from new EMMIX-WIRE (EM-W) model 
with $\theta_h^2$ equal to 1.3 and $d^2$ equal to 0) \label{Tab:4}}
{\begin{tabular}{lllllll}\toprule
&\multicolumn{2}{c}{First component}&\multicolumn{2}{c}{Second
component}
&\multicolumn{2}{c}{Third component}\\
\cline{2-7}
Parameters&EM-W&Kim&EM-W&Kim&EM-W&Kim\\\midrule
$p$(0.585,    &-0.001 & 0.024  &0.002  & -0.005 & -0.001  &-0.019 \\
0.1,0.315)  &(0.014)&(0.029) &(0.016) &(0.017) &(0.017) &(0.026)\\
$a_0$(0.3,    &-0.001  &0.018   &0.003  & -0.046 &0.000   &-0.005\\
1,0.2)      &(0.027)&(0.035) &(0.026) &(0.053) &(0.021) &(0.021)\\
$a_1$(0.03,   &0.001  &-0.068  &0.005   &-0.041  &0.001   &0.008\\
1,0.02)     &(0.085)&(0.146) &(0.108) &(0.127) &(0.086) &(0.085)\\
$b_1$(0.06,   &0.003  & -0.031&0.005  & 0.047  &0.002   &0.004 \\
0.9,0.01)   &(0.042)&(0.063) &(0.054) &(0.072) &(0.050) &(0.054)\\
\hline
$\theta^2$(1.3,& -0.059 & 1.254  & -0.076  &1.251    &-0.052   &1.242 \\
1.3,1.3)    &(0.087) &(1.254) & (0.178) & (1.257) &(0.104) &(1.243)\\
$\rho$(0.6    &0.012   &-0.198  & -0.013  & -0.201  &0.009   &-0.203 \\
0.6,0.6)    &(0.019) &(0.199) & (0.039) &(0.206)  &(0.023) &(0.204)\\
\hline
$\sigma^2$(1.0,& 0.046    & &0.056   && 0.039   &\\
1.0,1.0)       &(0.070)   & &(0.145) && (0.084) &\\
$d^2$(0.,      & 0.000    & & 0.001 && 0.000  &\\
0.,0.)         &(0.000)   & &(0.001) &&(0.000) &\\
\hline
& & EM-W & & & Kim & \\
\hline
Error rate & &0.154  & & & 0.162  & \\
Rand       & &0.796  & & & 0.783  & \\
Adjusted   & &0.590  & & & 0.566  & \\\botrule

\end{tabular}}{}
\end{table} 

We now compare our model with \cite{kim08} using the data from the 
extended model of \cite{qin06} , which is a special case of our
EMMIX-WIRE model (with $d^2$ = 0), where gene expressions are independent. 
The results are presented in Tables 3 and 4. 
As we explained in the last paragraph, the system errors are removed 
in this situation. And our model has unbiased estimation for all parameters. 
On the other hand, the model of \cite{kim08} still overestimates the 
residual variance at different time points and underestimates the 
correlation between different time points for each gene, 
as it fails to capture the contribution from 
gene-specific effects to the auto-correlation among periodic gene expressions
at each time point. 
Their method again produces larger error rates and slightly smaller 
Rand Indices.

Lastly, we generate the data from the model of \cite{kim08} 
and provide comparative results in Tables 5 and 6. 
It is observed from Tables 5 and 6 that the clustering performances 
are comparable between the two models. 

\begin{table}[!t] 
\processtable{Bias and standard deviation in brackets from 1000 simulated 
data points (generated from \cite{kim08} with $\theta_h^2$ equal to 0.5) 
\label{Tab:5}}
{\begin{tabular}{lllllll}\toprule
&\multicolumn{2}{c}{First component}&\multicolumn{2}{c}{Second
component}
&\multicolumn{2}{c}{Third component}\\
\cline{2-7}
Parameters&EM-W&Kim&EM-W&Kim&EM-W&Kim\\\midrule
$p$(0.585,    &-0.003 & 0.000   &-0.008  & 0.001 & 0.010  &-0.000 \\
0.1,0.315)  &(0.004)&(0.003) &(0.023) &(0.003) &(0.024) &(0.004)\\
$a_0$(0.3,  &0.002  &0.000   &0.003  & 0.001 &0.001   &0.001\\
1,0.2)      &(0.013)&(0.013) &(0.010) &(0.010) &(0.010) &(0.010)\\
$a_1$(0.03, &0.015  &0.001  &-0.236   &-0.002  &0.047   &0.003\\
1,0.02)     &(0.041)&(0.036) &(0.333) &(0.037) &(0.073) &(0.035)\\
$b_1$(0.06, &0.014  & -0.000 &-0.308  & -0.001  &0.058   &0.001 \\
0.9,0.01)   &(0.026)&(0.021) &(0.345) &(0.023) &(0.067) &(0.025)\\
\hline
$\theta^2$(0.5,&-0.034 & -0.000  & -0.006  &-0.001    &-0.021   &-0.000 \\
0.5,0.5)       &(0.036) &(0.006) & (0.027) & (0.015) &(0.025) &(0.009)\\
$\rho$(0.6     &0.020   &-0.000  & 0.013  & -0.001  &0.023   &-0.001 \\
0.6,0.6)       &(0.021) &(0.007) & (0.025) &(0.017)  &(0.028) &(0.009)\\
\hline
$\sigma^2$(0.0,&0.025      & &0.014   && 0.022   &\\
0.0,0.0)       &(0.026)   & &(0.015) && (0.023) &\\
$d^2$(0,       & 0.000   & & 0.045 && 0.042  &\\
0,0)           & (0.000)  & &(0.095) &&(0.056) &\\
\hline
& & EM-W & & & Kim & \\
\hline
Error rate & &0.018  & & & 0.016  & \\
Rand       & &0.978  & & & 0.980  & \\
Adjusted   & &0.955  & & & 0.960  & \\\botrule

\end{tabular}}{}
\end{table}

\begin{table}[!t] 
\processtable{Bias and standard deviation in brackets from 1000 simulated 
data points (generated from \cite{kim08} with $\theta_h^2$ equal to 1.3) 
\label{Tab:6}}
{\begin{tabular}{lllllll}\toprule
&\multicolumn{2}{c}{First component}&\multicolumn{2}{c}{Second
component}
&\multicolumn{2}{c}{Third component}\\
\cline{2-7}
Parameters&EM-W&Kim&EM-W&Kim&EM-W&Kim\\\midrule
$p$(0.585,    &-0.009 & 0.001  &-0.007  & 0.005 & 0.016  &-0.001 \\
0.1,0.315)  &(0.013)&(0.010) &(0.012) &(0.011) &(0.020) &(0.013)\\
$a_0$(0.3,    &-0.002  &-0.000   &0.015  & 0.001 &0.003   &-0.000\\
1,0.2)      &(0.023)&(0.023) &(0.024) &(0.019) &(0.016) &(0.016)\\
$a_1$(0.03,   &-0.005  &-0.001  &0.054   &-0.000  &0.003   &0.000\\
1,0.02)     &(0.071)&(0.074) &(0.0928) &(0.083) &(0.068) &(0.064)\\
$b_1$(0.06,   &0.015  & -0.000&-0.131  & 0.001  &0.020   &0.000 \\
0.9,0.01)   &(0.036)&(0.036) &(0.135) &(0.045) &(0.041) &(0.043)\\
\hline
$\theta^2$(1.3,& -0.195 & -0.000  & -0.185  &-0.003    &-0.186   &-0.002 \\
1.3,1.3)    &(0.196) &(0.016) & (0.192) & (0.049) &(0.189) &(0.025)\\
$\rho$(0.6    &0.043   &-0.000  & 0.037  & -0.002  &0.044  &-0.001 \\
0.6,0.6)    &(0.043) &(0.007) & (0.042) &(0.022)  &(0.045) &(0.010)\\
\hline
$\sigma^2$(0.0,& 0.144    & &0.131   && 0.143  &\\
0.0,0.0)       &(0.145)   & &(0.133) && (0.144) &\\
$d^2$(0.,      & 0.000    & & 0.000 && 0.001  &\\
0.,0.)         &(0.000)   & &(0.001) &&(0.001) &\\
\hline
& & EM-W & & & Kim & \\
\hline
Error rate & &0.101  & & & 0.102  & \\
Rand       & &0.864  & & & 0.866  & \\
Adjusted   & &0.725 & & & 0.729  & \\\botrule

\end{tabular}}{}
\end{table}

Our model again provides unbiased estimates for all parameters. 
In contrast to the model of \cite{kim08}, 
our model accounts for the correlation among gene profiles 
via the linear effects modelling. 
As presented in Tables 1 to 6, our model outperforms the model 
of \cite{kim08} when the genetic profiles are correlated.
When the genetic profiles are generated independently, our model has better
performance in cases where the variability in gene expressions at each time
point is large. In cases where the residual covariance structure follows an
AR(1) model (Kim et al., 2008), our model still provides comparative results 
and unbiased estimates as the model of \cite{kim08}. 
The advantage of our model is to provide more reliable and robust clustering of 
time-course data is apparent. With microarray experiments including those 
time-course studies, gene expression levels measured from the same tissue 
sample (or time point) are correlated \citep{gjm04}, 
clustering methods which assume independently distributed gene profiles, 
such as the model of \cite{kim08}, may overlook important sources of 
variability in the experiments, resulting in the consequent possibility 
of misleading inferences being made \citep{ng06}.

\subsection{Applications: Yeast cell cycle datasets}
 
\subsubsection{Yeast cell cycle dataset 1} The first data is the yeast cell cycle data with MIPS criterion from \cite{wong07}. 
This data set is extracted from \cite{cho01} and made available 
by \cite{yeung01}. 
The yeast cell cycle dataset contains 237 genes and 17 samples. 
These genes corresponding to four categories in the MIPS database 
(DNA synthesis and replication, organization of centrosome, nitrogen,
and sulphur metabolism, and ribosomal proteins); 
these are assumed to be the true clusters. 
In this illustration, we fit our new extended EMMIX-WIRE model 
and the model of \cite{kim08} 
to the yeast cell cycle data, 
with the period of 85 in the Fourier extension \citep{luan04}. 

\begin{figure}[]
\centering
\includegraphics[width=8cm,height=6cm]{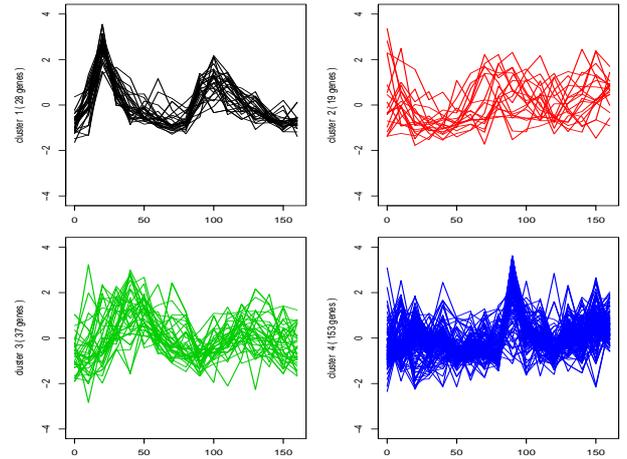}
\caption{Clustering of gene expression profiles into four groups for the yeast dataset 1.}
\label{fig:1}
\end{figure}

In the Table 2 of \cite{wong07}, it shows that the Rand and 
adjusted Rand Indices for their two-stage method are 0.7087 and 0.3697, 
respectively, and these indices are higher 
than other methods considered in their paper. 
Using the model of \cite{kim08}, 
the Rand indices are 0.7330 and 0.4721, respectively. 
With the model of EMMIX-WIRE \citep{ng06}, 
we have the Rand and adjusted Rand Indices 0.7799 and 0.5568, 
respectively. 
Using the proposed new model, 
the Rand and adjusted Rand Indices are 0.8123 and 0.6189, respectively, 
and are the best matches (the largest index) compared 
with the aforementioned models. 
The four clusters of genes time-course profiles 
are presented in Figure 1. 
It can be seen that the genes have very similar expression patterns 
within each cluster, except in cluster 2, 
where there is greater individual variation by some of the genes. 
The estimation using the proposed model is listed in Table 7. 
It can be seen that the correlations in the first three components 
are from 0.27 to 0.72, indicating a significant correlation 
among gene expressions at different time points. 
Ignoring this correlation may therefore lead to a lower Rand Index, 
that is, a worse clustering. 
We can see the estimates of $d^2$ in clusters 1 and 4 
are large and are greater than the corresponding estimates of $\theta^2$, 
indicating co-regulation in these two clusters. 
If we ignore such within-cluster co-regulation, 
we will have Rand Indices similar to those of \cite{kim08}. 
Our model considers both autocorrelation and co-regulation, 
and thus obtains the best clustering performance.  
  
\begin{table}[!t] 
\processtable{Estimations of parameters for the yeast cell cycle dataset 1 (237 genes) \label{Tab:7}}
{\begin{tabular}{lllll}\toprule
&first cluster&second cluster&  third cluster&fourth cluster\\\midrule
  $p$	     &0.104	&0.054	&0.118	&0.724\\
  $a_1$	     &-0.107&	0.400&	-0.807&	0.298\\
  $b_1$	     &1.009	&-0.119	&-0.053	&0.079\\
  $\sigma^2$ &	0.027	&0.011	&0.025	&0.278\\
  $\theta^2$	&0.174	&0.417	&0.443	&0.307\\
  $\rho$ &ĉ	0.278	&0.717	&0.435	&0.053\\
  $d^2$	     &0.191	&0.001	&0.031	&0.310\\
  $\omega$	 &85	&85	&85	&85\\\botrule

\end{tabular}}{}
\end{table} 

\begin{figure}[]
\centering
\includegraphics[width=8cm,height=8.6cm]{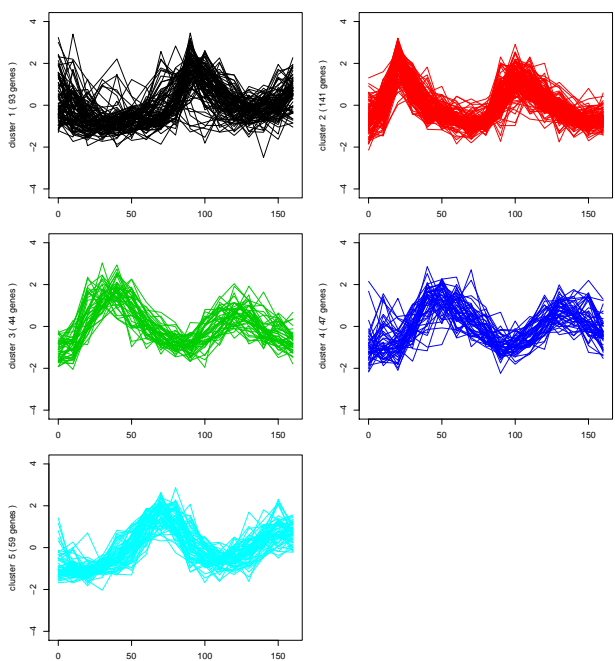}
\caption{Clustering of gene expression profiles into five groups for the yeast dataset 2.}
\label{fig:2}
\end{figure}

\begin{table}[!t] 
\processtable{Estimations of parameters for the yeast cell cycle dataset 2 (384 genes) \label{Tab:8}}
{\begin{tabular}{llllll}\toprule
&first cluster	&second	cluster&third cluster	&fourth	cluster &fifth cluster\\\midrule
  $p$	     &0.238	&0.290	&0.151	&0.165	&0.157\\
  $a_1$	     &0.643	&-0.061	&-0.736	&-0.616&	0.329\\
  $b_1$      &	-0.062	&1.019	&0.285	&-0.772	&-1.001\\
  $\sigma^2$ &0.011	&0.046	&0.037	&0.028	&0.006\\
  $\theta^2$ &0.498	&0.296	&0.470	&0.309	&0.244\\
  $\rho$ĉ	 & 0.503&	0.269&	0.364	&0.379&	0.550\\
  $d^2$	     &0.062	&0.052	&0.044	&0.065	&0.030\\
  $\omega$   &	85&	85	&85&	85&	85\\\botrule

\end{tabular}}{}
\end{table} 

\subsubsection{Yeast cell cycle dataset 2} The second example is the subset 
of 384 genes from the yeast cell cycle data \citep{cho01} while the full 
data set can be found from the Stanford yeast cell cycle web site (
\verb+http://171.65.26.52/yeast_cell_cycle/+ \\
\verb+cellcycle.html+).

Each of gene is assigned a "phase". We call each "phase" a "Main Group". 
There are five "Main Groups" in this dataset, 
namely, early G1, late G1, S, G2 and M. 
We now compare and assess the cluster quality with the 
external criterion (the 5 phases). 
The raw data is log transformed and normalized by columns and rows. 
Figure 2 presents the five clusters of genes profiles obtained 
using the proposed model. 
It can be seen that the genes have very similar expression patterns 
within each cluster. 
The estimations are listed in Table 8. 
The Rand and adjusted Rand Indices are 0.8102 and 0.4484, respectively. 
They are 0.8108 and 0.4592 for the model of \cite{kim08}. 
The error rates are the same (0.2813) for the two models. 
The performances of the two models are very similar 
because the correlation among gene profiles is weak in this dataset. 
As indicated in Table 8, the estimates of $d^2$ 
are all very small compared to the estimates of $\theta^2$. 
 
\section{DISCUSSION}

We have presented a new mixture model with AR(1) random effects 
for the clustering of time-course gene expression profiles. 
Our new model involves three elements taking important role in modelling 
time-course periodic expression data, namely, 
(a) Fourier expansion which models the periodic patterns; 
(b) auto-correlation variance structure that accounts 
for the auto-correlation among the observations at different time points; 
and (c) the cluster-specific random effects which incorporate 
the co-regulation within the clusters. 
In particular, the latter two elements corresponding to the 
correlations between time-points and between genes are crucial 
for reliable and accurate clustering of time-course data. 
We have demonstrated in the simulation and real examples that the 
accuracy of clustering is improved if the auto-correlation 
among the time dependent gene expression profiles has been accounted 
for along the time points; this is also demonstrated in \cite{kim08}. 
Furthermore, better results are obtained if the co-regulation 
within the clusters is modelled appropriately. 
When the correlation between genetic profiles is not small, 
which is the case for typical time-course data, ignorance of this 
dependency may lead to less accurate clustering results. 

For the purpose of comparison, the periods of the signal of gene expression 
are assumed to be known in the simulation study and applications to 
real data. 
In practice, there are several ways to estimate the periods 
for each cluster (\citealp{kim08}; \citealp{luan04}; \citealp{spellman98}; 
\citealp{ng06}). 
For example, in \cite{kim08}, the periods are estimated using 
simplex algorithm at the M-step during the EM algorithm. 
However, when the periods are estimated during the EM iterations, 
we find that the periods depend also on other parameters. 
In addition, when we start from an initial period and get the 
design matrix X, then with higher possibility the best period 
will be the initial periods. 
So we change the strategy to a slow one, 
and we call it global grid search method, 
which guarantees the highest maximum log likelihood at the best periods. 
It performs as follow, let S is a set with its element 
as (period $\omega_1$, period $\omega_2$, $\dots$, 
period $\omega_g$), where $\omega_h$ can take all possible 
values (grid points). 
For example, for the yeast cell cycle data, 
the possible periods are 60, 61, $\dots$, 90. 
Then for each fixed ($\omega_1$, $\omega_2$,$\dots$, $\omega_g$), 
we estimate the parameters as if the periods for each component are known. 
Finally we compare the log likelihood and choose the one 
with the highest log likelihood as the final result. 
Since it is very slow if there are too many elements in set S 
when we have no prior information about periods, 
we recommend use other method to get the periods first \citep{booth08}. 
In all the calculation in this paper, 
we assume the period is fixed, that is, 
there is only one element in the set S. 

The proposed model is very flexible through the different specifications 
of design matrices or model options as originally available in \cite{ng06}. 
For example, besides the full model, 
it enables us to incorporate the model of \cite{qin06} 
as a special case. 
Specifically, we can obtain their model by assuming zero cluster effects 
($v$ = 0) and that random effects $u$ be auto correlated for each gene. 
Furthermore, when both random effects $u$ and $v$ are assumed to be zero, 
then we have normal mixture of regression models. 
In the program we have developed, 
there are many options and parameters for users to specify the models 
they want to use in addition to the models we list in our paper. 
The program is written in R package and is available from the 
corresponding author.

\section*{Acknowledgement}
This work was supported by a grant from the Australian Research Council.

\noindent 
Funding: The Australian Research Council Discovery Project Grant (DP0772887).

\end{document}